\newcommand{\CLASSINPUTtoptextmargin}{0.9in}

\documentclass[10pt,  twocolumn, twoside, journal, final]{IEEEtran}

\usepackage{graphicx}
\usepackage{cite}
\usepackage{graphicx}
\usepackage{psfrag}
\usepackage{subfigure}
\usepackage{url}
\usepackage{stfloats}
\usepackage{amsmath}
\usepackage{array}
\usepackage{epsfig}
\usepackage{amssymb}
\usepackage{algorithm}
\usepackage{algcompatible}

\usepackage{xcolor}
\usepackage{amsthm}

\usepackage{mathtools} 

\algnewcommand\algorithmicreturn{\textbf{return}}
\algnewcommand\RETURN{\State \algorithmicreturn}%

\newcommand{\x}{\mathbf{x}}

\newcommand{\W}{\mathbf{W}}

\begin{document}

\title{Rate Analysis and Deep Neural Network Detectors for SEFDM FTN Systems}

\author{Arsenia Chorti,~\IEEEmembership{Senior Member,~IEEE}, David Picard,~\IEEEmembership{Member,~IEEE}
\tiny\thanks{ Arsenia Chorti is  with ETIS UMR8051 CY Cergy Paris Université, ENSEA, CNRS, France and David Picard is with the  École des Ponts ParisTech.}}\normalsize

\maketitle

\begin{abstract}
In this work we compare the capacity and achievable rate of uncoded faster than Nyquist (FTN)  signalling  in  the  frequency  domain,  also referred to as spectrally efficient FDM (SEFDM). We propose a deep residual convolutional neural network detector for SEFDM signals in additive white Gaussian noise channels, that allows to approach the Mazo limit in systems with up to $60$ subcarriers. Notably, the deep detectors achieve a loss less than $0.4-0.7$ dB for uncoded QPSK SEFDM systems of $12$ to $60$ subcarriers at a $15\%$ spectral compression.

\vspace{0.1 cm}
\textit{Index Terms:} Spectally efficient FDM (SEFDM), faster than Nyquist (FTN), convolutional neural networks, spectral efficiency.

\end{abstract}

\section{Introduction}

The celebrated Nyquist limit for the transmission rate guarantees aliasing-free reception, i.e., interference-free observations, which is the cornerstone of a simple digital signal processing (DSP) based detector. However, unlike the Shannon capacity, \textit{it is not a fundamental limit}, in the sense that if the samples reach the receiver at a rate faster than the Nyquist (FTN), there may not be a fundamental loss in performance associated with it \cite{Rusek,CTN17}, up to a certain limit; this so-called \textit{Mazo limit} has been empirically established to be around $25\%$ of the nominal rate in binary modulated signals \cite{Mazo75, Rusek05}. This increase in the transmission rate comes at the cost however of a substantial increase in transmitter complexity \cite{Chorti} when reducing interference levels using filter banks \cite{Fettweis, Osaki20}, or at the receiver when using advanced detection techniques as an alternative to the optimal vectorial maximum likelihood detection (MLD) \cite{Kanaras1, Kanaras2}.

In this work, we focus on a special class of FTN signals referred to as spectrally efficient frequency division multiplexing (SEFDM), in which the orthogonality of orthogonal FDM (OFDM) subcarriers is violated by reducing the intercarrier spacing to a fraction $\alpha \in (0, 1)$ of the nominal value.
The reason for focusing on SEFDM in particular, lies in the fact that unlike other proposed approaches in the time domain \cite{Rusek09} or using raised cosine filter banks to mitigate the interference levels \cite{Osaki20}, SEFDM can be implemented using simply an inverse fast Fourier transform (IFFT) \cite{Izzat}. The fact that SEFDM can be implemented on standard DSP is a major advantage, rendering SEFDM a frontrunner FTN waveform among the many possibilities \cite{Banelli, FBMC}, well-suited for applications with constrained transmitters. As in SEFDM the complexity burden is moved to the receiver, it is better suited for the uplink rather than the downlink; e.g., for  massive machine type communications (mMTC) in which the bandwidth crunch is more accentuated, SEFDM can provide a possible alternative to non-orthogonal multiple access (NOMA) that requires coordination between the users \cite{Bello}.

In this contribution we investigate two aspects of SEFDM systems; first, the potential capacity and rate gains with respect to standard OFDM, and, secondly, the implementation of the SEFDM optimal MLD detector using deep residual convolutional neural networks (CNNs). Notably, we readily demonstrate that SEFDM can offer important gains in terms of spectral efficiency with respect to OFDM and furthermore we confirm Mazo's limit for SEFDM systems with up to \textcolor{black}{60} quadrature phase shift keying (QPSK) modulated SEFDM subcarriers. This work also opens up the discussion for understanding the limits in structured interference cancellation using deep CNN detectors.

\section{SEFDM Capacity}
Let us consider a SEFDM system with $N$ non-orthogonal FDM subcarriers at a normalised subcarrier spacing denoted by $\alpha \in (0,1)$; as an example, for $\alpha=0.9$ the  subcarrier spacing is reduced to $90\%$ of its nominal OFDM value.
Denoting the transmitted symbols vector by $\mathbf{s}=[s_0, s_1, \ldots, s_{N-1}]^T$ and the SEFDM subcarrier matrix by $\mathbf{F}^{\alpha}$, the  SEFDM signal can be expressed as
\begin{math}
    \mathbf{F}^{\alpha}\mathbf{s}, 
\end{math}
with
\begin{equation}
    [\mathbf{F}^{\alpha}]_{m,n}=\frac{1}{\sqrt{N}}\exp\left(2 \pi i \alpha \frac{mn}{N} \right), m,n=0,\ldots, N-1.
    \end{equation}
     Using the QR decomposition, we can decompose the subcarriers matrix as $\mathbf{F}^{\alpha}=\mathbf{Q} \mathbf{R}$, with $\mathbf{Q}$ an orthonormal matrix that corresponds to an orthonormal basis that spans the SEFDM signal space and $\mathbf{R}$ the upper triangular matrix that captures the projections of $\mathbf{F}^{\alpha}$ on $\mathbf{Q}^H$.  
In \cite{Chorti} the modified Gram Schmidt procedure was proposed to perform the aforementioned QR decomposition in a numerically stable manner.

Assuming an additive white Gaussian (AWGN) channel, at the SEFDM receiver the  incoming signal is projected onto $\mathbf{Q}^H$; note that contrary to simply using a matched filter receiver as proposed in \cite{Fettweis}, this receiver design preserves the whiteness and the Gaussianity of the noise. As a result, a vectorial MLD can be employed as an equivalent to the optimal maximum a posteriori detector. Based on the above discussion, the observed vector $\mathbf{y}=[y_0, \ldots, y_{N-1}]$ at the input of the MLD is expressed as
\begin{equation}
    \mathbf{y}=\mathbf{Q}^H\mathbf{F}^{\alpha}\mathbf{s}+\mathbf{Q}^H\mathbf{n}=\mathbf{R}\mathbf{s}
+\mathbf{{n}}_q,
\end{equation}
where $\mathbf{n, {n}}_q\sim \mathcal{CN}(\mathbf{0}, N_0\mathbf{I}_N)$ and $\mathbf{I}_N$ the $N\times N$ identity matrix.

Setting $\mathbf{R}=\mathbf{U \Sigma V}^H $ the singular value decomposition of the projections matrix $\mathbf{R}$ and $\mathbf{x}=\mathbf{V}^H\mathbf{s}$, we have that
\begin{equation}
\mathbf{\tilde{y}}=\mathbf{U}^H\mathbf{y}=\mathbf{U}^H\left( \mathbf{U\Sigma V}^H\mathbf{V x}+ \mathbf{n}_q\right) =\mathbf{\Sigma x + \tilde{n}} \label{eq:SVD}
\end{equation}
where $\mathbf{\tilde{n}}=\mathbf{U}^H\mathbf{n}_q \sim \mathcal {CN}(\mathbf{0}, N_0\mathbf{I}_N)$ since $\mathbf{U}^H$ is unitary. Having decomposed the observation vector of the SEFDM signal in a parallel stream of $N$ independent sub-channels as in (\ref{eq:SVD}), the capacity of the SEFDM system can be reached by using Gaussian signalling and the waterfilling algorithm to optimally allocate the available power over the parallel sub-channels, i.e., the optimal power allocation is given by
\begin{equation}
    p_i=\left(\mu- \frac{1}{\sigma_i^2} \right)^+, \text{ and }\mu\text{ chosen s.t. } \sum_{i=1}^N p_i=NP,
\end{equation}
while  $\mathbf{p}=[p_0, \ldots, p_M, 0, \ldots, 0]^T$ is the optimal power allocation, $\sigma_i$ denotes the $i-$th diagonal element of $\mathbf{\Sigma}$ and $P$ the average power budget per subcarrier. To achieve the SEFDM capacity we transmit the  SEFDM signal $\mathbf{F}^{\alpha}diag(\mathbf{p})\mathbf{V}^H\mathbf{s}$, using Gaussian codebooks. 

Notice that using the waterfilling power allocation on the symbol $\mathbf{x}=\mathbf{V}^H \mathbf{s}$ corresponds to not transmitting the last $N-M$ SEFDM subcarriers as these are assigned zero power. At the same time, interestingly, the information on the last $N-M$ symbols of $\mathbf{s}$ is still ``conveyed'' thought the first $M$ SEFDM subcarriers through the precoding matrix $\mathbf{V}^H$. 

From the precious discussion, the SEFDM capacity (i.e., spectral efficiency in bits/sec/Hz) is given by
\begin{equation}
    C_{\text{SEFDM}}= \frac{1}{\alpha N}\sum_{i=1}^{M}{\log_2\left(1+\frac{\sigma_i^2 p_i}{N_0}\right)}.
\end{equation}
If on the other hand equal power allocation (i.e., no precoding) is used among all SEFDM subcarriers, then the achievable rate of the \textit{uncoded} SEFDM is given by 
\begin{equation}
    R_{\text{SEFDM}}=\frac{1}{\alpha N} \sum_{i=1}^{N}{\log_2\left(1+ \frac{\sigma_i^2 P}{N_0} \right)}.
\end{equation}


\section{SEFDM Detection Using Deep Neural Networks}
The detection of SEFDM signals can be approached with deep neural networks as a multi-class classification problem. In particular, assuming an $M$-QAM SEFDM system, the observation vector $\mathbf{y}$ can be used to infer the ``class'' $c(s_i)$ of the transmitted symbols $s_i, i=0, \ldots, N-1$. In the case of OFDM the projections matrix $\mathbf{R}$ is the identity matrix and as a result the classification problem is easy, i.e., the corresponding Voronoi regions of the optimal classifier are the well known $M$-QAM MLD regions. 

In the case in which $\mathbf{R}$ is no longer diagonal for $\alpha<1$, the Voronoï partition on $\mathbf{s}$ is deformed in a non trivial way and standard MLD can no longer recover the class $c(s_i)$ from $\mathbf{y}$. We thus want to learn a prediction function $f$ whose goal is to predict the class $c(s_i), i=0, \ldots,N-1$ of the transmitted symbols given the observation vector $\mathbf{y}$, i.e.,
\begin{align}
    \min _f \sum_{\mathbf{s}} \mathbb{E}_\mathbf{w}[ l(f(\mathbf{y}), c(\mathbf{s}))],
\end{align}
where $l(\cdot, \cdot)$ denotes a cost function measuring the error between $f(\mathbf{s})$ and $c(\mathbf{s})$. 
As the joint probability distribution of $(f(\mathbf{y}, c(\mathbf{s}))$ is unknown, we can use supervised machine learning to obtain $f$, i.e., 
a neural network of parametric functions $h_l: \mathbb{R}^{2N} \rightarrow \mathbb{R}^{w_l}$ (instead of complex numbers, we manipulate 2 dimensional real vectors), followed by a projection $g: \mathbb{R}^D \rightarrow \mathbb{R}^M$ onto the classes simplex.

Each function is called a \emph{layer}. A neural network is characterized by 3 parameters: the family of functions $h$ that are used, its depth $d$ which corresponds to the number of layers and its width $w_l$ which corresponds to the size of the intermediate space which the inputs are mapped to. Note that since $h$ are not linear, it makes sense to increase the dimension ($w \geq 2N$) to \emph{unfold} the transformation that was performed by $\mathbf{R}$.
The gradient descent is performed over all parameters of all layers using the chain rule. 

In this paper, we propose to use CNNs with residual connections. The building block of each layer $l$ is a convolution with a weight matrix, followed by the non linear activation:
\begin{align}
    \x_l &= h_l(\mathbf{x_{l-1}}) =  [(\W_{il} \star \x_{l-1})_+]_{i\leq w_l},\\
    \x_0 &= \mathbf{y},
\end{align}
with $\W_{il} \in \mathbb{R}^{k\times d}$ is the weight matrix, $k$ corresponds to the window size (\emph{kernel size}) of the convolution and $(t)_+ = \max(0, t)$ is the non linear activation function (ReLU). The output of each layer is a stack of several  such convolutions, hence the name CNNs.

At first, it may seem counter-intuitive to use convolutions as $\mathbf{y}$ is by no mean translation invariant (or shift invariant) which is their main appeal. Here, we use convolutions as a structural regularization for $f$ to reduce the risk of overfitting.
We tried simpler architectures using fully connected layers (multiple layer perceptron - MLP), but we found experimentally that the problem requires significant depth that made MLP overfit due to their high number of parameters.
In contrast, convolutions are naturally sparse predictor as each dimension $i$ in the output $\x_{l+1}$ is only seeing a subset of the dimensions of the input $\x_l$. With sufficient depth however, all dimensions of $\mathbf{y}$ are combined as the cumulative kernel size gets over $2N$.

Furthermore, residual connections are added to ease the training process by avoiding that the gradient vanishes after being back-propagated through too many layers. This amounts to skipping every two layers:
\begin{align}
\label{eq:res}    x_{2l+2} &= h_{2l+2}(\mathbf{x}_{2l+1}) + \mathbf{x}_{2l}\\
\label{eq:res2}    x_{2l+1} &= h_{2l+1}(\mathbf{x}_{2l})
\end{align}
Residual connections have been proven very popular in computer vision applications~\cite{he2016cvpr}, where the need for training very deep neural networks emerges from the highly non-linear nature of the problems considered.

\section{Numerical Results}
In this section we present two sets of results. First, we investigate the capacity and rate gains when using SEFDM with $\alpha \in \{0.9, 0.85,  0.8\}$ as opposed to OFDM versus the signal to noise ratio (SNR) and the number of subcarriers $N$. Secondly, we demonstrate that using deep CNN detectors,  Mazo's empirical observation for no degradation in performance can be approached  in uncoded SEFDM systems of up to $\textcolor{black}{N=60}$ subcarriers, which to the best of our knowledge is the largest SEFDM system for which this is observed.

\subsection{SEFDM Capacity and Rates}

\begin{figure}[t]
     \centering
     \includegraphics[
     width=\columnwidth]{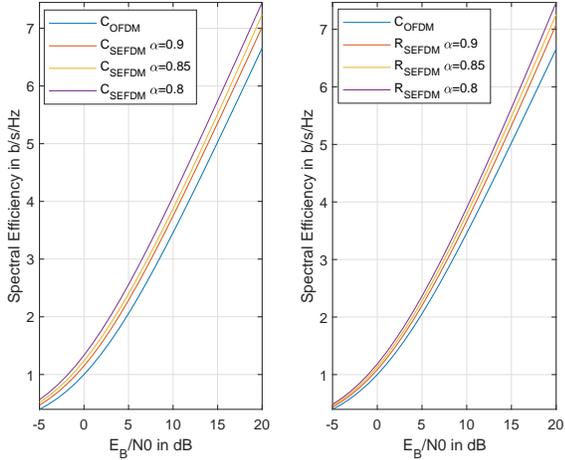}
     \caption{Capacity and rate comparison of SEFDM and OFDM for $N=12$, vs the energy per bit to the noise spectral density.}
     \label{fig:N12}
 \end{figure}
 
\begin{figure}[t]
     \centering
     \includegraphics[
     width=\columnwidth]{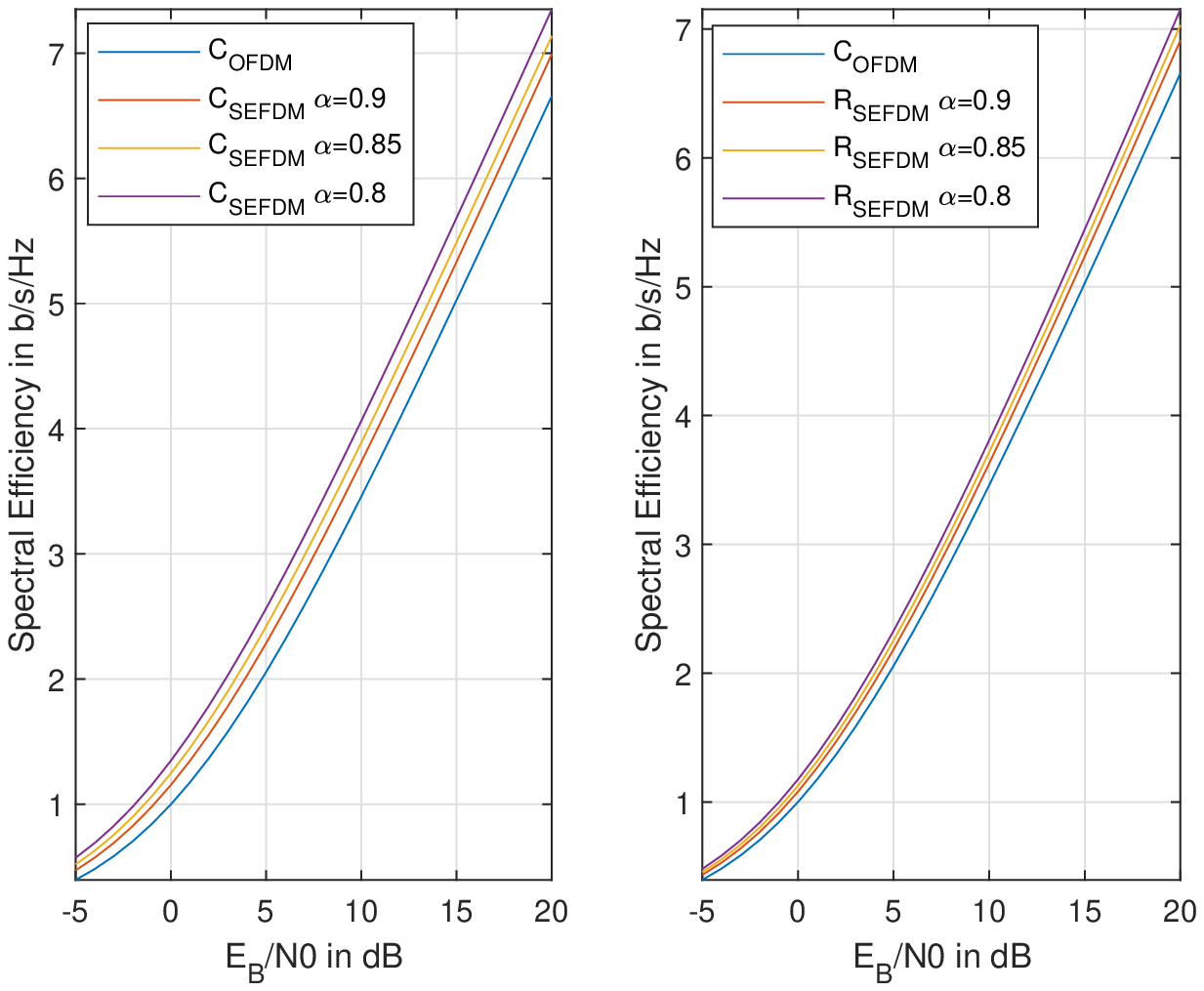}
     \caption{Capacity and rate comparison of SEFDM and OFDM for $N=48$, vs the energy per bit to the noise spectral density.}
     \label{fig:N48}
 \end{figure}
 \begin{figure}[t]
     \centering
     \includegraphics[
     width=\columnwidth]{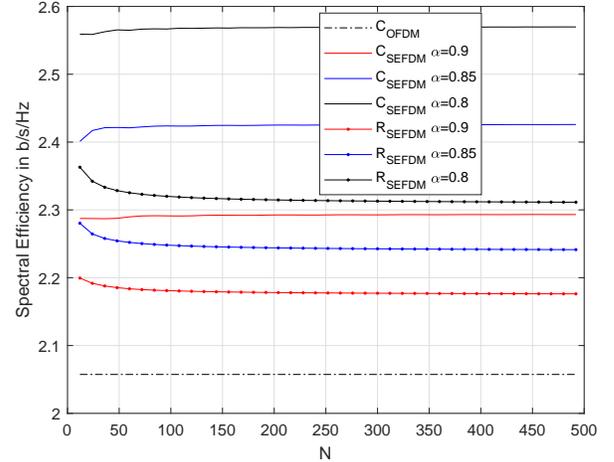}
     \caption{Capacity and rate comparison between SEFDM and OFDM for $N=48$}
     \label{fig:20dB}
 \end{figure}
\begin{figure}[t]
    \centering
     \includegraphics[
     width=\columnwidth]{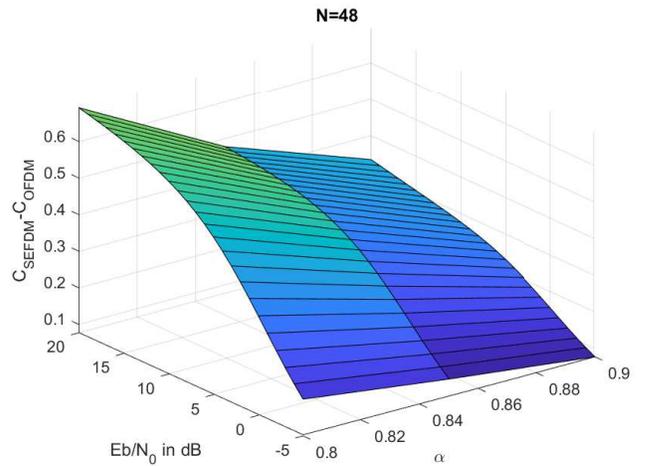}
     \caption{Capacity gap between SEFD and OFDM  versus the SNR and $\alpha$ for $N=42$ subcarriers.}
     \label{fig:3D_N12}
 \end{figure}
\begin{figure}[t]
     \centering
     \includegraphics[
     width=\columnwidth]{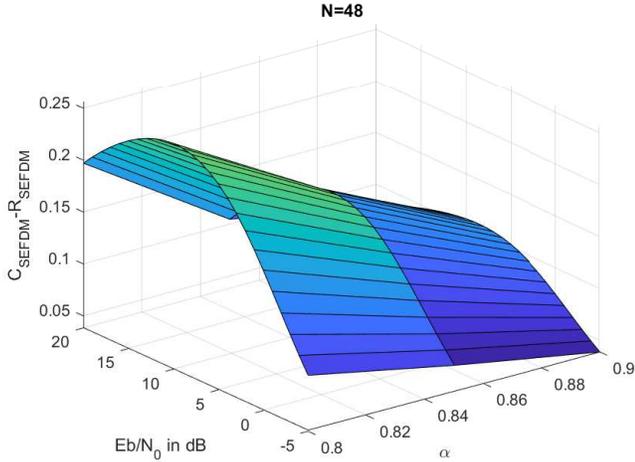}
     \caption{Rate gap between the SEFD capacity and uncoded SEFDM rate  versus the SNR and $\alpha$ for $N=42$ subcarriers}
     \label{fig:2D_N48}
 \end{figure}

In fifth generation (5G) systems a physical resource block (PRB) comprises $N=12$ subcarriers, so in the following we present results for SEFDM systems with $N$ a multiple of a PRB. In Figs. \ref{fig:N12} and \ref{fig:N48}, the rates achieved when using the SEFDM with waterfilling power allocation, $C_{\text{SEFDM}}$, or with equal power across all subcarriers, $R_{\text{SEFDM}}$, are shown versus the $E_b/N_0$ for $\alpha \in \{0.9, 0.85, 08\}$ for $N=12$ and $N=48$, respectively. The SEFDM offers gains in terms of spectral efficiency even in the case of uncoded SEFDM. 

Furthermore, in Fig. \ref{fig:20dB} we focus in the high SNR region $E_b/N_0=20$ dB and demonstrate that the SEFDM capacity is independent of $N$ and only varies with $\alpha$. Interestingly, for uncoded SEFDM, the achievable rate is slightly higher for $N<50$. The related rate gaps are further depicted over the whole range of SNR and $\alpha$ values in Figs. \ref{fig:3D_N12} and \ref{fig:2D_N48} for $N=48$. Notice that the rate gap between the SEFDM capacity and the uncoded SEFDM is lower in the low and high SNR regions while it increases with $\alpha$.
 


\subsection{Deep Residual CNN Detector}
Next, we demonstrate that the Mazo limit can be approached when using deep residual CNN detectors. In particular, we show that for up to 60 subcarriers and for $\alpha \leq 0.85$  the performance of the uncoded SEFDM (i.e., with equal power allocation) is preserved as conjectured by Mazo. In other words, we show that the bit error rate (BER) of the uncoded SEFDM is roughly the same as that of a baseline OFDM when QPSK modulation is used. To the best of our knowledge, it is the first time this has been reported in the literature for $N>40$ \cite{KanarasLetter}, especially in the low SNR region. 

We train the neural networks by performing $10^5$ gradient descent steps over 256 randomly generated observations $\mathbf{y}$ at each step, using Adam optimizer with a learning rate of 0.001 and the categorical entropy loss as cost function.
The network is then evaluated on $10^5$ randomly generated observations to compute the bit error rate (BER).
The training observations are generated with a fixed SNR of $E_B/N_0 = 0$ dB, whereas observations used to evaluate the network once it is trained are made with varying SNR. Among all the architectures we tried, we found that deep residual CNNs had the best compromise between computational complexity and attained error rate.
The best architecture is composed of three scales of increasing width $w\in \{24, 48, 96\}$, each composed of 3 blocks of residual convolutions as described in (\ref{eq:res}) and (\ref{eq:res2}). Accounting for the convolutions to increase the width before each scale and the final layer that performs the classification, this architecture has $3\times (3\times 2 +1) +1 = 22$ layers in total.
As is usually observed in deep learning, we found that depth was more beneficial than width.

The results depicted in Figs. 6 to 10 for $N=\{ 12, 24, 36, 48, 60 \}$ demonstrate that for $\alpha\leq 0.85$ the BER of uncoded SEFDM follows very closely that of a baseline QPSK system with standard detection. Remarkably, the loss is less than $0.4$ dB at an $Eb/N_0$ of 7dB for $N=12$ and less than $0.7 $ dB for $N=60$ subcarriers, even though the CNNs were trained to maximize performance at 0 dB. To the best of our knowledge it is the first instance where so small  BER losses are confirmed for SEFDM FTN systems with of than $40$ subcarriers. In future work the deep detectors will be tested in optimally precoded SEFDM systems using the waterfilling approach outlined in Section II, trained to optimize performance at higher SNRs.


 \begin{figure}[t]
     \centering
     \includegraphics[width=\columnwidth]{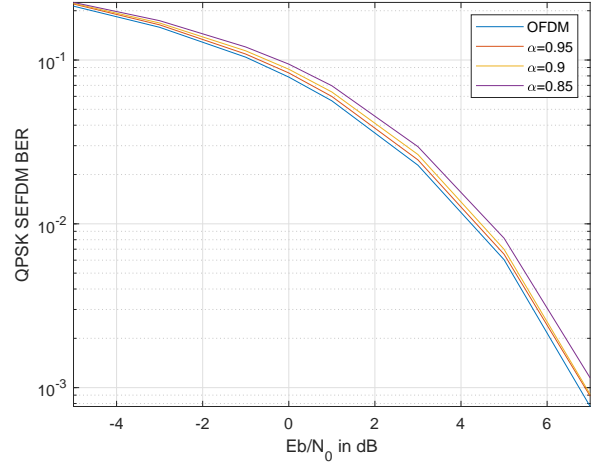}
     \caption{Comparison of BER of uncoded SEFDM with OFDM for QPSK modulation and $N=12$}
     \label{fig:alphaN12}
 \end{figure}

 \begin{figure}[t]
     \centering
     \includegraphics[width=\columnwidth]{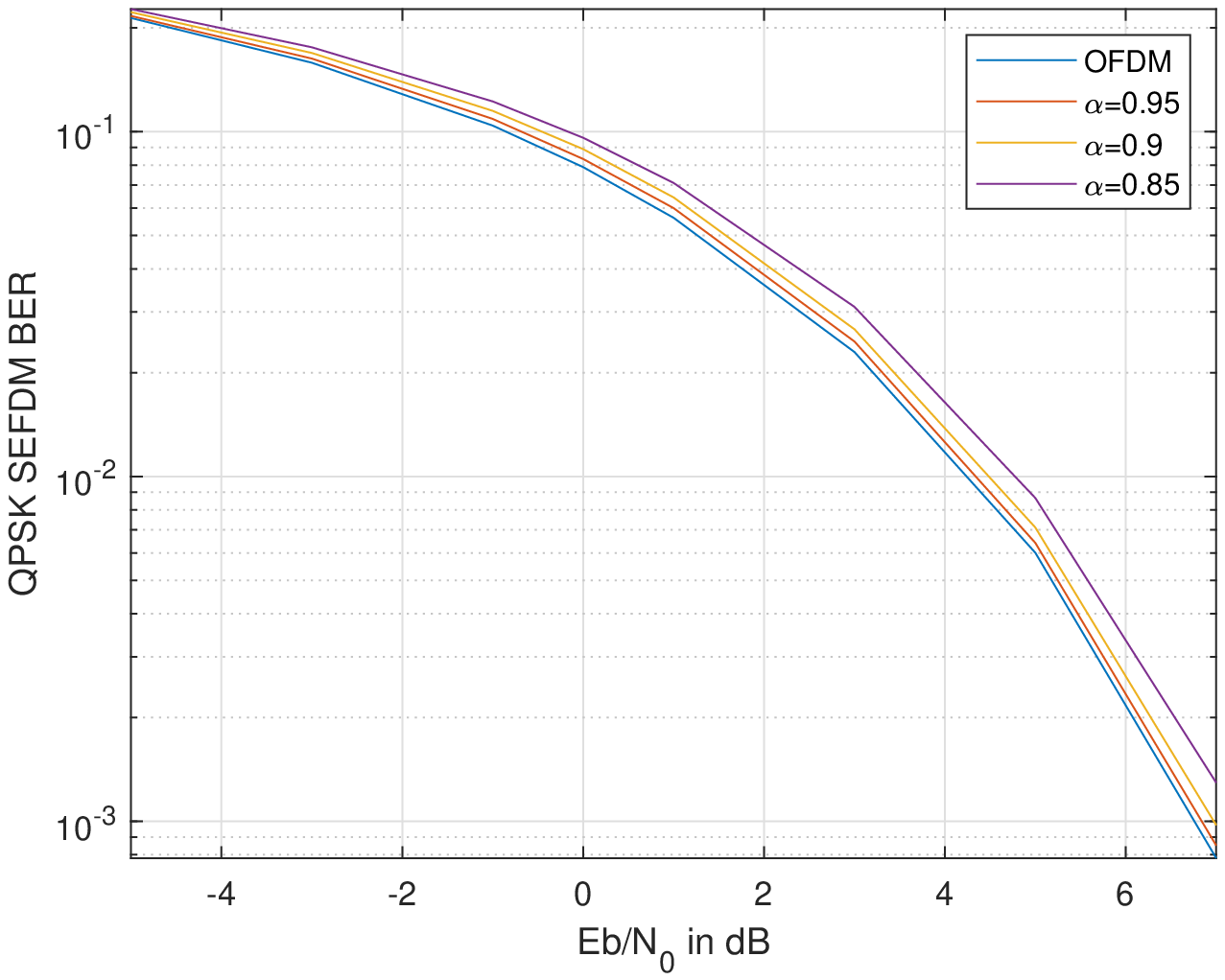}
     \caption{Comparison of BER of uncoded SEFDM with OFDM for QPSK modulation and $N=24$}
     \label{fig:alphaN24}
 \end{figure}

 \begin{figure}[t]
     \centering
     \includegraphics[width=\columnwidth]{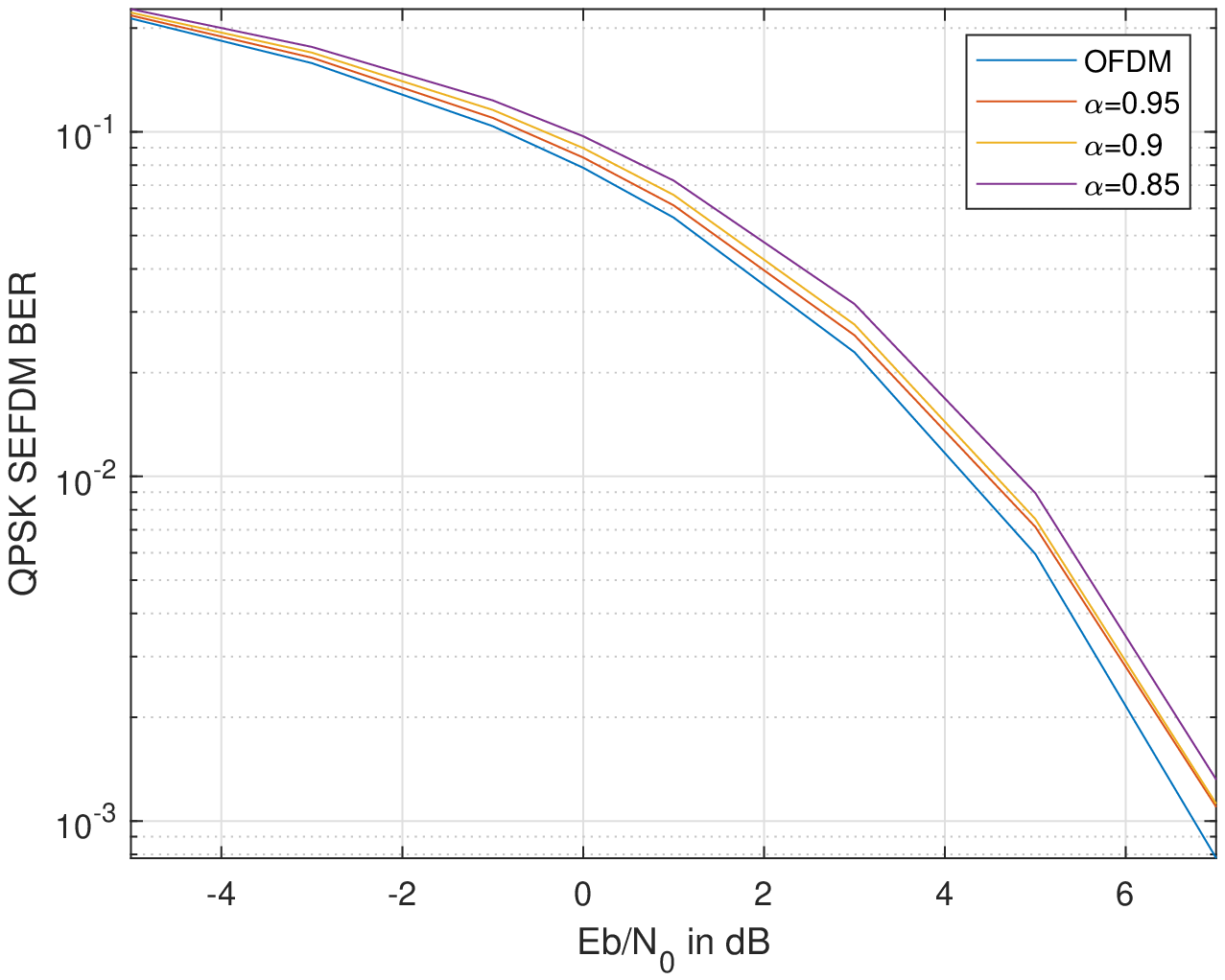}
     \caption{Comparison of BER of uncoded SEFDM with OFDM for QPSK modulation and $N=36$}
     \label{fig:alphaN36}
 \end{figure}

 \begin{figure}[t]
     \centering
     \includegraphics[width=\columnwidth]{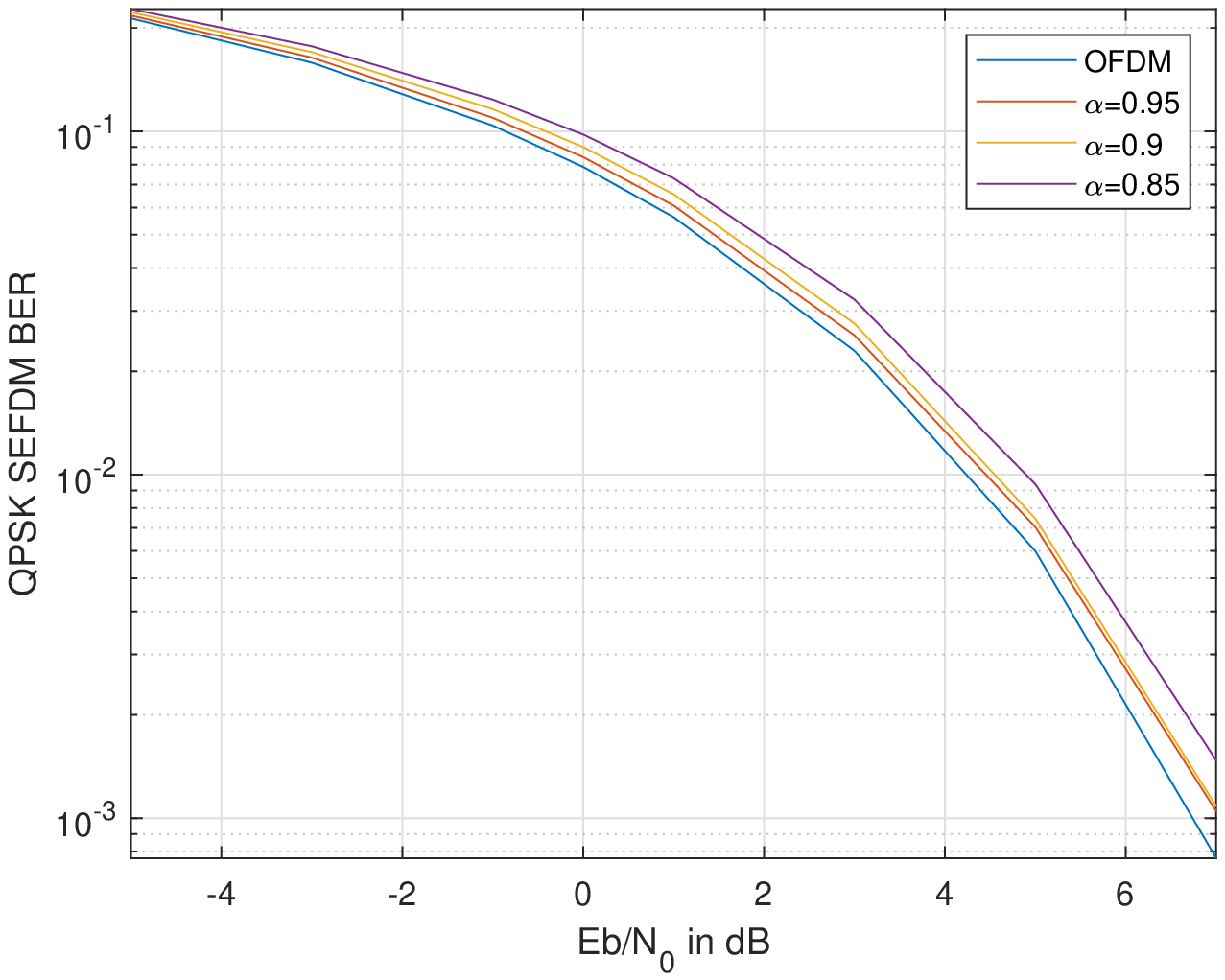}
     \caption{Comparison of BER of uncoded SEFDM with OFDM for QPSK modulation and $N=48$}
     \label{fig:alphaN48}
 \end{figure}

 \begin{figure}[t]
     \centering
     \includegraphics[width=\columnwidth]{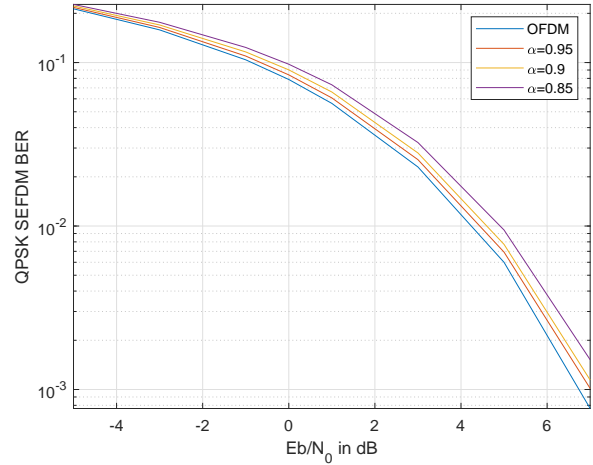}
     \caption{Comparison of BER of uncoded SEFDM with OFDM for QPSK modulation and $N=60$.}
     \label{fig:alphaN56}
 \end{figure}

These results demonstrate that residual CNNs can be employed as deep learning based detectors in systems with high structured interference levels and confirm that it is possible to approach the Mazo limit even for uncoded SEFDM systems.

\section{Conclusions}
In this letter the capacity and achievable rates of SEFDM systems were investigated and SEFDM was shown to bear significant gains with respect to standard OFDM in terms of spectral efficiency. Furthermore, a novel deep residual CNN based detector was proposed and shown to approach the Mazo limit for systems of up to $60$ SEFDM uncoded subcarriers. These results show that SEFDM could be considered for the uplink of bandwidth constrained systems such as mMTC without compromising performance.
\bibliographystyle{IEEEtran}

\bibliography{main.bib}
\end{document}